\documentclass{article}

\usepackage{booktabs}
\usepackage{longtable}
\usepackage{tabularx}
\newcolumntype{R}{>{\raggedleft\arraybackslash}X}%
\usepackage{subfig}
\usepackage[table]{xcolor} \definecolor{highlightcol}{gray}{0.85}
\usepackage[%
  textsize=tiny,%
  backgroundcolor=white,%
  bordercolor=gray,%
  linecolor=gray]%
  {todonotes}
\usepackage[utf8]{inputenc}
\usepackage[T1]{fontenc}

\usepackage[english]{babel}
\usepackage{ifthen}
\usepackage{xifthen}
\usepackage{amsmath}
\usepackage{amsfonts}
\usepackage{amssymb}
\usepackage{marvosym}
\usepackage{mathtools}
\usepackage[inline]{enumitem}
\newlist{deepitemize}{enumerate}{5}
\setlist[deepitemize]{leftmargin=*,label=--}
\setenumerate[1]{label=(\arabic*)}
\usepackage{multicol}
\usepackage{float}
\usepackage{tikz}
\usetikzlibrary{arrows, shadows}
\usepackage{gnuplot-lua-tikz}
\usepackage{listings}
\PassOptionsToPackage{hyphens}{url} 
\usepackage[bookmarks=true]{hyperref}
\hypersetup{%
  bookmarksdepth=3,%
  pdfborder={0 0 0},
}

\usepackage{epsfig}
\usepackage{multirow}
\usepackage{caption}
\usepackage{graphicx}
\usepackage[ruled,vlined,linesnumbered]{algorithm2e}
\SetKwBlock{ExecutionEnvironment}{Execution environment:}{}
\SetKwBlock{Verifier}{Verifier:}{}
\SetKwBlock{Initialization}{Initialization:}{}
\SetKwBlock{InstrumentProgram}{Instrument program:}{}
\SetKwBlock{BeforeMA}{Before-memory-access:}{}
\SetKwBlock{AfterMA}{After-memory-access:}{}
\SetKwBlock{SchedulerRequest}{Scheduler on receiving request of thread TID:}{}
\SetKwBlock{SchedulerNotification}{Scheduler on receiving completion notification of thread TID:}{}
\usepackage{cite}
\usepackage{amsthm}
\newtheorem{definition}{Definition}
\newtheorem{theorem}{Theorem}
\newtheorem{example}{Example}
\usepackage{supertabular}

\newcommand{\mathnotation}[3]{%
  \expandafter\newcommand\csname #1\endcsname[#2]{\ensuremath{#3}}%
}
\newcommand{\mn}[3]{\mathnotation{#1}{#2}{#3}}

\newcommand{\zeroargnot}[2]{%
  \expandafter\newcommand\csname #1\endcsname{}%
  \expandafter\def\csname #1\endcsname/{#2}%
}

\newcommand{\zeroargnotmath}[2]{%
  \zeroargnot{#1}{\ensuremath{#2}}%
}
\newcommand{\zam}[2]{\zeroargnotmath{#1}{#2}}
\newcommand{\zeroargnotmathsingleletter}[2]{%
  \zeroargnotmath{x#1}{#2}
}
\newcommand{\zamsl}[2]{\zeroargnotmathsingleletter{#1}{#2}}

\zamsl{naturals}{\mathbb{N}}

\mn{powerset}{1}{\mathcal{P}(#1)}

\zam{concatenate}{\cdot}

\zamsl{P}{P}

\zamsl{V}{V}

\zamsl{E}{E}

\mn{states}{1}{S_{#1}}

\zamsl{s}{s}

\zam{initstates}{S_{\mathit{init}}}

\mn{sinit}{1}{s_{#1}}

\mn{events}{1}{E_{#1}}

\mn{constraints}{1}{C_{#1}}

\zam{predicates}{\mathcal{F}}

\mn{remove}{1}{\mathit{remove}(#1)}

\zamsl{c}{c}

\mn{traces}{1}{\mathit{traces}(#1)}

\mn{executions}{1}{\mathit{executions}(#1)}

\zamsl{o}{o}

\zamsl{O}{O}

\zam{traceprefix}{\preccurlyeq}

\zam{tracematch}{\approx}

\mn{transitions}{1}{\rightarrow_{#1}}

\mn{transition}{2}{\xrightarrow{#1}_{#2}}

\zamsl{t}{t}

\zamsl{e}{e}

\mn{alphabet}{1}{\Sigma_{#1}}

\mn{tid}{1}{\mathit{tid}(#1)}

\mn{en}{1}{\mathit{enabled}(#1)}

\mn{exec}{2}{(#1, #2)}
\zamsl{u}{u}
\zamsl{v}{v}
\zamsl{w}{w}

\mn{complete}{1}{\Downarrow_{#1}}

\zamsl{U}{U}

\zamsl{emptysequence}{\varepsilon}

\zamsl{emptyset}{\varnothing}

\mn{free}{1}{\mathit{free}(#1)}

\zamsl{G}{G}

\mn{safe}{1}{\mathit{safe}_{\xG/}(#1)}

\mn{errorfree}{1}{\mathit{error\_free}(#1)}

\zam{finished}{\mathit{finished_{\xG/}}}

\zam{admissibleprefix}{\xo/_{\mathit{adm}}}

\begin{document}

\title{\Large \bf Safe Execution of Concurrent Programs by Enforcement
  of
  Scheduling Constraints \\[1.5em]
  \small \normalfont Patrick Metzler, Habib Saissi, P\'eter Bokor, and
  Neeraj
  Suri \\[1em]
  Technische Univerist\"at Darmstadt \\
  \{metzler, saissi, pbokor, suri\}@deeds.informatik.tu-darmstadt.de}
\date{}
\maketitle

\begin{abstract}
  Automated software verification of concurrent programs is challenging
because of exponentially large state spaces with respect to the number
of threads and number of events per thread. Verification techniques
such as model checking need to explore a large number of possible
executions that are possible under a non-deterministic
scheduler. State space reduction techniques such as partial order
reduction simplify the verification problem, however, the reduced
state space may still be exponentially large and intractable.

This paper discusses \emph{Iteratively Relaxed Scheduling}, a
framework that uses scheduling constraints in order to simplify the
verification problem and enable automated verification of programs
which could not be handled with fully non-deterministic
scheduling. Program executions are safe as long as the same scheduling
constraints are enforced under which the program has been verified,
e.g., by instrumenting a program with additional synchronization. As
strict enforcement of scheduling constraints may induce a high
execution time overhead, we present optimizations over a naive
solution that reduce this overhead. Our evaluation of a prototype
implementation on well-known benchmark programs shows the effect of
scheduling constraints on the execution time overhead and how this
overhead can be reduced by relaxing and choosing constraints.


\end{abstract}

\section{Introduction}
Concurrent programs with non-deterministic scheduling may show
exponentially large state spaces with respect to the number of threads
and events per thread, which is a hurdle for automated verification
techniques such as model checking~\cite{Valmari1996}. In contrast to
testing, verification covers all possible program behaviors, including
different behaviors due to non-deterministic scheduling. In order to
reduce the complexity of the verification task, model checking
techniques such as partial order reduction (POR)~\cite{Clarke1999,
  Godefroid1996, Flanagan2005} exist that reduce a concurrent
program's state space. Although recent POR
algorithms~\cite{Wachter2013, Abdulla2014, Albert2017, Sousa2017} show
considerable improvements, even reduced state spaces may be of
exponential size~\cite{Godefroid1996} and state of the art tools for
automated verification may need a long time to verify a program. In
other cases, state of the art tools may even not terminate at all for
a given program. Even if it is possible to manually rewrite such a
program to make it tractable by a particular tool, such a process
would be time consuming. Overall, the delay that is introduced by
verification between completion of the implementation of a concurrent
program and deployment, the \emph{verification delay}, is presumably
still unacceptably high for a wide industrial adoption of
verification.

\emph{Iteratively Relaxed Scheduling} (IRS)~\cite{Metzler2017} is a
framework for verification of safety properties on concurrent programs
under non-deterministic scheduling. It enables a reduction and
adjustment of the verification delay by enforcing scheduling
constraints that ensure a safe program execution even when only a
fraction of the state space is already proven to be safe. By
iteratively using intermediate verification results, scheduling
constraints can be relaxed to gradually increase the amount of
non-determinism of scheduling.

Permitting a safe execution of a program before complete verification
has finished is a key novelty of IRS and distinguishes it from other
verification techniques that use intermediate verification results. In
IRS, it is necessary to restrict scheduling of programs in order to
ensure that only correct executions may occur. We are aware of only
two classes of approaches to restricting scheduling: (A) Deterministic
multi-threading (DMT) and related techniques facilitate concurrency
testing by reducing the amount of non-determinism due to concurrency,
thereby reducing the number of necessary test cases and improving
reproducibility of errors. Scheduling of concurrent programs is
restricted such that scheduling is deterministic for a particular
input or only a reduced set of schedules may occur for a particular
input~\cite{Liu2011, Cui2013}. (B) Another approach to reduce
non-determinism is to synthesize synchronization statements, which
allows to restrict scheduling independently of inputs, in contrast to
(A). Representatives of this class are automated fence insertion,
which has been applied to facilitate verification of concurrent
programs with relaxed memory models~\cite{Burckhardt2008, Fang2003}
and synchronization synthesis such as~\cite{Gupta2015}, which inserts
locks and other synchronization primitives that are more powerful than
fences in that also scheduler-related non-determinism can be
eliminated.  Methods in (A) depend on concrete inputs and are
therefore not suited for verification, as a change in the program
inputs may have an effect on scheduling that is unforeseeable for the
verifier. The effectiveness of (B) is limited as only non-determinism
due to relaxed memory accesses is removed (in case of fence insertion)
or deadlocks may be introduced~\cite{Gupta2015}, which limits the
program's functionality.

Compared to such existing approaches, IRS shows several differences
that make it a candidate for more effective reduction of the
verification delay. IRS uses intermediate verification results to
record and iteratively increase knowledge about schedules that adhere
to a given program specification (we denote such schedules as
\emph{correct}). Such schedules are guaranteed to show correct
behavior independently of program inputs. IRS applies scheduling
constraints in order to enforce that only correct schedules occur. As
more schedules are explored during verification and thus known to be
correct, scheduling constraints can be relaxed. In contrast to (A),
determinism can be enforced independently of program inputs and the
amount of non-determinism is controllable dynamically (during program
execution) via scheduling constraints. Additionally, unlike some
techniques in (A), IRS is able to provide strong
determinism~\cite{Olszewski2009}. In contrast to (B), IRS addresses
scheduler-related non-determinism and enables to adjust the amount of
non-determinism between fully deterministic and fully
non-deterministic executions.

Common to all the aforementioned techniques is execution time overhead
caused by additional synchronization. A central assumption underlying
the feasibility of IRS is that scheduling constraints can be designed
such that the execution time overhead drops below an acceptable
threshold. This paper validates this assumption for different classes
of benchmark programs. Hence, IRS allows us to find a sweet spot
between a low verification overhead and a low execution time overhead.

In order to iteratively relax scheduling constraints, it is necessary
to collect knowledge about correct schedules already during the
verification process rather than monolithically at the end. In
contrast to existing model checking approaches that use intermediate
verification results, IRS comprises additional requirements: an
intermediate verification result proves safety under some scheduling
constraints. These constraints must permit to fully use the program,
i.e., no restrictions on the program inputs or the execution length
may be applied.

This paper investigates the enforcement of scheduling constraints in
IRS. A formal interface between IRS and a model checker is
presented. Based on scheduling constraints that satisfy the
requirements of this interface, the execution time overhead of
constrained executions with respect to unconstrained executions is
analyzed. For a presentation of iterative model checking that can
generate scheduling constraints, we refer to~\cite{Metzler2019}.

\textbf{Contributions.} We provide a formal framework and algorithm
for IRS that (1) states requirements for verification algorithms on
their intermediate results and (2) enables a correct execution of a
program provided with such an intermediate verification result,
without introducing deadlocks. (3) We evaluate the effect of relaxing
scheduling constraints on execution time overhead, based on an
optimized scheduling implementation of IRS. Our results show that
execution time overhead depends not only on the number of scheduling
constraints but on their structure as well. Our implementation and
experimental data is publicly available.


\section{Overview}\label{sec:naive-irs}
\emph{Iteratively relaxed scheduling} (IRS) is an iterative
verification approach for concurrent programs with non-deterministic
scheduling, in contrast to the conventional, monolithic approach to
program verification. This section reviews the basic IRS approach that
we have previously presented~\cite{Metzler2017}. Detailed definitions
are provided in Section~\ref{sec:optimization}. The conventional
approach to program verification can be described as follows:
\begin{enumerate*}
\item Develop a program or update an existing program.
\item Verify the (updated) program.
\item Repeat steps (1)--(2) until the verification is successful.
\item The (updated) program can be safely used under a
  non-deterministic scheduler, i.e., with all feasible schedules.
\end{enumerate*}
Since all feasible schedules have to be verified, the verification
step may be slow or may even not terminate, in which case a different
verification technique or tool would have to be tried or the program
would have to be rewritten such that it is tractable by the used
verification technique. In any of these cases, a large verification
delay is introduced.

By constraining the scheduler and thereby reducing the number of
feasible schedules, IRS tries to reduce the verification delay. A
program can be safely used as soon as a sufficiently large initial set
of safe schedules is found. An \emph{IRS execution environment}
ensures that only safe schedules are feasible. Safe schedules that are
found later can be used to iteratively relax the scheduling
constraints. In particular, the verification approach of IRS is:
\begin{enumerate*}
\item Develop a program or update an existing program.
\item Start an iterative verification process which verifies in each
  iteration an individual schedule or a set of schedules.
\item As soon as a sufficient subset of safe schedules is found, the
  program can be safely used inside an IRS execution environment.
\item New safe schedules that are found in subsequent iterations may
  relax the scheduling constraints during execution of the program.
\end{enumerate*}

The set of initially found safe schedules must enable a program
execution without restricting functionality. Hence, for each possible
input, a safe schedule must be available. Additionally, a specific use
case may require additional initially safe schedules, for example to
allow a practical enforcement of scheduling constraints.

Initial experiments have shown that constraining scheduling may
introduce considerable execution time
overhead~\cite{Metzler2017}. Only if it is possible to reduce the
execution time overhead by relaxing scheduling constraints, the
overhead incurred by IRS can be adjusted: the more schedules are
verified, the less overhead will occur. In this case, the sweet spot
between a short verification delay and a small execution time overhead
can be found by continuously testing the execution time overhead with
the current set of schedules found to be safe. As soon as the
execution time overhead is small enough (i.e., a ``sufficient amount
of non-determinism'' is used), the program can be used and
verification can be stopped (i.e., no more than the ``necessary amount
of non-determinism'' is used).

If a program shows a sufficient set of schedules to use it without
restricting functionality and additionally shows schedules that
violate the specification, IRS can be used to nevertheless safely
execute the program. The program may be fixed, in which case it may be
possible to eventually remove all scheduling constraints, or it may be
left unchanged and used only with scheduling constraints that
guarantee safety.

Several conceivable use cases are given in~\cite{Metzler2017}, including
\begin{enumerate*}
\item safely use programs that are not tractable by conventional verification,
\item safely use programs with concurrency bugs,
\item verify as many schedules as possible within a given time budget,
  and
\item verify as many schedules as necessary for a given budget of execution
  time performance.
\end{enumerate*}

An IRS execution environment may be realized inside an application
program or by modifying the operating system. For example, in the
former case, the program may be instrumented so that a thread waits
before memory accesses that are not yet permitted to occur, according
to the scheduling constraints. Even if the scheduler of the operating
system is non-deterministic, the scheduling constraints are
enforced. In the latter case, it is conceivable to directly constrain
the scheduler of the operating system to obtain an IRS execution
environment and enforce schedules.

Unlike previous approaches to deterministically execute concurrent
programs (e.g.,~\cite{Liu2011, Cui2013, Cui2011}), IRS provides a
novel approach to constraint scheduling independent from program
inputs. This independence makes it compatible with program
verification. However, only such verification techniques are suitable
for IRS that yield meaningful intermediate verification
results. Meaningful intermediate verification results either show a
counter example for program correctness or guarantee correctness under
some feasible scheduling constraints. No additional constraints should be
necessary such as constraints about program inputs or execution
length, as a program may not be fully operational under such
constraints. In particular, a correct schedule has to be known for
each possible program input, even if inputs are given interactively
(during a program execution).

Intermediate verification results of this form have previously not
been proposed, to the best of our knowledge. Nevertheless, some
existing verification approaches show similarities or potential to be
applied in the framework of IRS. Forward search-based, symbolic model
checking, such as the Impact algorithm for concurrent programs by
Wachter, Kroening, and Ouaknine~\cite{Wachter2013}, seems to be
suitable to generate intermediate verification results for IRS, by
using a depth-first strategy. This strategy might use heuristics that
prioritize executions according to their number of context switches,
as it is done in context bounding analyses, e.g., by Qadeer and
Rehof~\cite{Qadeer2005} or in \emph{iterative context bounding} by
Musuvathi and Qadeer~\cite{Musuvathi2007}. Furthermore, verification
in IRS can partition the state space of a program into disjoint sets
of schedules and verify each partition individually. A similar
approach has been followed for bug finding by Nguyen et
al.~\cite{Nguyen2017}. Conditional model checking by Beyer et
al.~\cite{Beyer2012a} could be used in the framework of IRS if
conditions passed between verification runs encode suitable
intermediate results.

\emph{Mazurkiewicz traces} are equivalence classes on program
executions and are used in the context of partial order reduction
(POR)~\cite{Guo2015, Wachter2013, Flanagan2005, Gueta2007,
  Abdulla2014} to identify executions that can be skipped during
verification. In order to formalize suitable intermediate verification
results, IRS uses an extended notion of Mazurkiewicz traces, called
\emph{symbolic traces}. Based on symbolic traces, we formalize the
requirements on a verifier for IRS in the following section.

As in the related field of DMT, additional synchronization is
necessary to enforce scheduling constraints. Our experiments confirm
that constraining scheduling introduces a considerable execution time
overhead, in extreme cases a 44 times slowdown. A main concern for the
praticality of IRS is to limit this overhead depending on the
requirements of a use case. We try to design IRS with a low overhead
by addressing several aspects: the amount of additional
synchronization for schedule enforcement, storage and look-up of
scheduling constraints, and the effect of relaxing constraints on the
execution time overhead.

Section~\ref{sec:optimization} addresses the former aspects: reducing
the amount of additional synchronization is achieved by permitting
threads to execute as long as possible without interruption. For an
efficient encoding scheduling constraints, we introduce \emph{trace
  prefixes}, based on symbolic traces. Section~\ref{sec:evaluation}
investigates the latter aspect of execution time overhead due to
scheduling constraints.


\section{The IRS Algorithm}\label{sec:optimization}
An IRS algorithm consists of a \emph{verifier} and an \emph{execution
  environment}, which run concurrently. The verifier continuously
verifies schedules and reports sets of safe schedules to the execution
environment. In order to guarantee program executions without
functional restrictions, submitted sets of safe schedules must
represent feasible schedules for all program inputs. This restriction
for suitable intermediate verification results are formalized below as
\emph{symbolic traces}. A symbolic trace permits at least one schedule
for each possible program input. A symbolic trace that is reported by
the verifier to the execution environment is called an
\emph{admissible trace}. The execution environment maintains
scheduling constraints, which are updated for each submitted
admissible trace.

When implementing IRS, it is desirable to efficiently maintain and
enforce scheduling constraints in order to incur as little overhead as
feasible over conventional program execution. This section presents a
detailed IRS algorithm and discusses its efficiency with respect to:
\begin{itemize}
\item Storing scheduling constraints
\item Looking up scheduling constraints
\item Synchronization between threads for the enforcement of
  scheduling constraints
\end{itemize}

\begin{algorithm}[t]
\caption{General IRS}
\label{alg:basic}
\KwData{\xV/ -- the set of admissible traces, initially empty}

\Verifier{
  \For{each trace \xo/ in \traces{\xP/}}{
    verify \xo/\;
    \If{\xo/ is safe}{
      add \xo/ to the set of admissible traces \xV/\;
    }
  }
}

\ExecutionEnvironment{
  set the current execution (\sinit{0}, \xu/) to the empty sequence\;
  \While{\xP/ has not terminated}{
    choose some thread \xt/ s.t.\ \((\sinit{0}, \xu/ \concatenate/
      \xt/) \traceprefix/ \xo/\) for some admissible trace \(\xo/ \in \xV/\)\;\label{line:basic:check}
    execute the next event of \xt/\;
    append \xt/ to \xu/\;
  }
}
\end{algorithm}

Algorithm~\ref{alg:basic} shows the general IRS
algorithm~\cite{Metzler2017}. It consists of a \emph{verifier} and an
\emph{execution environment}, which run concurrently. The verifier
maintains a set of admissible traces \xV/, which is initially
empty. Each time a trace has been verified and found to be safe, it is
added to \xV/. The execution environment starts to execute the program
in a given initial state \sinit{0} and records the execution as
\((\sinit{0}, \xu/)\).
Before each concurrent event, it is checked whether the current
execution \((\sinit{0}, \xu/)\)
adheres to some admissible trace in \xV/
(line~\ref{line:basic:check}). If \xV/ does not yet contain such a
trace, the execution is blocked until the verifier has found a trace.

As more and more schedules or symbolic traces are proven to be
correct, they are added to the set of admissible traces. This
representation of scheduling constraints has an exponential space
requirement and it seems impractical to store all symbolic traces for
large programs. Similarly, when permission for an event is checked,
the look-up time is exponential if no further structure is given to
the set of admissible traces. \emph{Unfoldings} have been applied for
model checking both Petri nets~\cite{McMillan1992} and concurrent
programs~\cite{Kahkonen2012, Rodriguez2015, Sousa2017}. By unfoldings,
it is possible to represent all executions of a concurrent program in
a single data structure, which is more space-efficient than storing a
set of all symbolic traces since each event occurs only once in an
unfolding. Looking up an event in an unfolding is faster than
searching in an unstructured set of symbolic traces, as well. However,
the size of an unfolding can still grow quickly (exponentially in the
worst case) with an increasing number of
threads~\cite{Kahkonen2014}. The space efficiency of verification
based on a depth-first search is lost. Hence, unfoldings are not
directly suitable to store scheduling constraints for practical
programs. In order to implement IRS, we address the problem of space
complexity by using \emph{trace prefixes}. If all admissible
Mazurkiewicz traces or executions are stored in order to express
scheduling constraints, so that each time a new execution has been
verified and is permitted, more space is required. In contrast, trace
prefixes can be used as scheduling constraints such that when new
executions are permitted, constraints may be removed and \emph{less}
space is required. However, the use of trace prefixes requires the
verifier to explore symbolic traces in a depth-first manner. More
freedom can be given to the verifier by extending trace prefixes to
partial unfoldings, at the price of a higher space
requirement. Additionally, an interesting question, however left for
future work, is how to generalize scheduling constraints for
non-terminating programs, for example by representing scheduling
constraints via cyclic graphs or automata.

Our tests of several IRS implementations confirmed that as expected,
inter-thread synchronization incurs a major part of execution time
overhead of IRS over unconstrained scheduling.  In order to reduce the
execution time overhead caused by synchronization between threads, it
is crucial to omit such synchronization in case an event needs not to
be scheduled after an event from an other thread. The IRS algorithm
presented in this section achieves this by executing several events
without intermediate synchronization, as is detailed below. Besides
reducing the amount of inter-thread synchronization, execution time
overhead can be considerably reduced by reducing the duration of a
single synchronization, for example by using lock-free synchronization
instead of locks. We discuss this matter in
Section~\ref{sec:implementation}.

In the following, we state our system model and present the IRS
algorithm, proving correctness and progress of the algorithm.

\subsection{System Model}
We model a (concurrent) \emph{program} \xP/ as a transition system
\((\states{}, \initstates/, \alphabet{}, \transitions{})\)
where \states{} is a set of states,
\(\initstates/ \subseteq \states{}\)
is a set of initial states (program inputs), \alphabet{} is a finite
set of threads, and
\(\transitions{} \subseteq (\states{} \times \alphabet{})
\rightharpoonup \states{}\)
is an acyclic transition relation. We require that for a given state
and thread, there is at most one successor state, i.e., scheduling is
the only source of non-determinism. We write
\(\xs/_1 \transition{\xt/}{} \xs/_2\)
to denote \((\xs/_1, \xt/, \xs/_2) \in \transitions{}\).

An \emph{execution} of \xP/ is a state and a sequence of \emph{events}
\((\sinit{0}, \xu/) \in \states{} \times (\alphabet{} \times
\xnaturals/)^*\),
where \(\xu/ = \xe/_1 \dotsc \xe/_n\),
such that the following holds.  (1) There exist states and transitions
such that
\(\sinit{0} \transition{\xt/_1}{} \xs/_1 \dotsb \transition{\xt/_n}{}
\xs/_n\)
and (2) each event \(\xe/_i = (\xt/_i, k)\)
contains thread \(\xt/_i\)
and counts occurrences of \(\xt/_i\)
in \xu/ before position \(i\):
\(k = |\{\xe/_j : j < i \wedge \xe/_j = (\xt/_i, \_)\}|\).
We denote the thread \xt/ of an event \(\xe/ = (\xt/, k)\)
by \tid{\xe/}. We write \(\sinit{0} \transition{\xu/}{} \xs/_n\)
if execution \((\sinit{0}, \xu/)\)
leads to state~\(\xs/_n\).
A state \xs/ is \emph{terminal} if no thread \xt/ and no state
\(\xs/'\)
exists such that \(\xs/ \transition{\xt/}{} \xs/'\).
An execution \((\sinit{0}, \xu/)\)
is complete, written \complete{(\sinit{0}, \xu/)}, if it leads to a
terminal state, otherwise, it is partial. The set of all executions of
\xP/ is denoted by \executions{\xP/}.

We assume that there exists a set \predicates/ of state
predicates. For example, each state may contain a valuation that maps
program variables to integer values; state predicates are encoded as
first order formulas over program variables and symbols for integer
arithmetic. For a predicate \(\xc/ \in \predicates/\),
we write \(\xs/ \vDash \xc/\)
(\(\xs/ \nvDash \xc/\)) if \xs/ satisfies (does not satisfy) \xc/.

We assume that a \emph{dependency} relation is given for every program
\xP/ that gives rise to a Mazurkiewicz equivalence relation on
executions~\cite{Mazurkiewicz1986, Abdulla2014} via a happens-before
relation between events. We extend the notion of Mazurkiewicz traces
by \emph{symbolic traces}, defined below. The happens-before relation
of one or more executions is represented by a \emph{symbolic trace
  graph} as a tuple \(\xo/ = (\events{\xo/}, \transitions{\xo/})\)
such that
\begin{itemize}
\item \(\events{\xo/} \subseteq (\alphabet{} \times \xnaturals/\)
  is a set of events and
\item
  \(\transitions{\xo/} \subseteq \events{\xo/} \times \predicates/
  \times \events{\xo/}\)
  is a partial order labeled with state predicates, which expresses a
  happens-before relation.
\end{itemize}

As an auxiliary function, we introduce \remove{\xe/, \xo/}, which
removes event \xe/ from symbolic trace graph \xo/. Formally,
\(\remove{\xe/, (\events{\xo/}, \transitions{\xo/})} =
(\events{\xo/}', \transitions{\xo/}')\) such that
\begin{itemize}
\item \(\events{\xo/}' = \events{\xo/} \setminus \xe/\) and
\item
  \(\transitions{\xo/}' = \{(\xe/_1, \xc/, \xe/_2) \in
  \transitions{\xo/} : \xe/_1 \neq \xe/ \wedge \xe/_2 \neq \xe/\}\).
\end{itemize}

In order to use symbolic trace graphs as scheduling constraints, we
introduce a notion to express that an execution satisfies the
scheduling constraints of a symbolic trace graph: an execution
\((\sinit{0}, \xu/)\)
with \(\xu/ = \xe/_1 \dotsc \xe/_n\)
and
\(\sinit{0} \transition{\tid{\xe/_1}}{} \xs/_1 \dotsc
\transition{\tid{\xe/_n}}{} \xs/_n\)
\emph{adheres} to the happens-before relation of a symbolic trace
graph \(\xo/ = (\events{\xo/}, \transitions{\xo/})\),
written \((\sinit{0}, \xu/) \traceprefix/ \xo/\), if \xu/ is empty, or
\begin{itemize}
\item \(\xe/_1 \in \events{\xo/}\),
\item
  \(\forall (\xe/, \xc/, \xe/') \in \transitions{\xo/} \ldotp \xe/' =
  \xe/_1 \Rightarrow \sinit{0} \nvDash \xc/\) and
\item
  \((\xs/_1, \xe/_2 \dotsc \xe/_n) \traceprefix/ \remove{\xe/_1,
    \xo/}\).
\end{itemize}
If \((\sinit{0}, \xu/) \traceprefix/ \xo/\)
and additionally, after removing all events from \xu/ from
\events{\xo/}, no events remain that need to be considered at
\(\xs/_n\),
we write \((\sinit{0}, \xu/) \tracematch/ \xo/\).
Formally, let
\((\events{n}, \transitions{n}) = \remove{\xe/_n, \remove{\xe/_{n-1},
    \dotsc \remove{\xe/_1, \xo/}}} \).
We define \((\sinit{0}, \xu/) \tracematch/ \xo/\)
as \((\sinit{0}, \xu/) \traceprefix/ \xo/\)
and
\(\forall \xe/ \in \events{n} \ldotp \forall (\xe/, \xc/, \xe/') \in
\transitions{n} \ldotp \xs/_n \nvDash \xc/\)
and call execution \((\sinit{0}, \xu/)\)
a \emph{linearization} of \xo/.

Based on symbolic trace graphs and their correspondence to
happens-before relations of executions, we define \emph{(symbolic)
  traces}, as a generalization of Mazurkiewicz traces. Intuitively, a
symbolic trace contains scheduling information for all possible
program inputs and represents all executions of a program with
matching scheduling.

\begin{definition}
  A \emph{(symbolic) trace} is a symbolic trace graph \xo/ such that
  \(\forall \sinit{0} \in \initstates/ \ldotp \exists \xu/ \ldotp
  (\sinit{0}, \xu/) \in \executions{\xP/} \wedge \complete{(\sinit{0},
    \xu/)} \wedge (\sinit{0}, \xu/) \tracematch/ \xo/\).
\end{definition}
A \emph{trace prefix} \xo/ is a trace, except that we do not require
executions to be complete:
\(\forall \sinit{0} \in \initstates/ \ldotp \exists \xu/ \ldotp
(\sinit{0}, \xu/) \in \executions{\xP/} \wedge (\sinit{0}, \xu/)
\traceprefix/ \xo/\).

\begin{figure}
  \centering
  \begin{minipage}{0.9\linewidth}
    \begin{multicols}{2}
\begin{lstlisting}
§\(T_1\)§:
    input: int *x
    local: int a
    a := *x
    *x := a + 1
    assert *x == a + 1§\columnbreak§
§\(T_2\)§:
    input: int *y
    local: int b
    b := *y
    *y := b + 1
\end{lstlisting}
    \end{multicols}
  \end{minipage}
  \caption{Example program.}
  \label{fig:example-program}
\end{figure}

\begin{figure}
  \centering
  \begin{tikzpicture}[->,>=stealth',shorten >=1pt,auto,node distance=1.8cm,
  thick,main node/.style={circle,draw}]

  \node[main node] (11) {};
  \node[left=0.1cm of 11] {\((T_1, 0):\) read x};
  \node[main node] (12) [below of=11] {};
  \node[left=0.1cm of 12] {\((T_1, 1):\) write x};
  \node[main node] (13) [below of=12] {};
  \node[left=0.1cm of 13] {\((T_1, 2):\) read x};

  \path[every node/.style={font=\sffamily\small,
        fill=white,inner sep=1pt}]
    (11) edge (12)
    (12) edge (13);

  \node[main node] (21) [right=3cm of 11] {};
  \node[right=0.1cm of 21] {\((T_2, 0):\) read y};
  \node[main node] (22) [below of=21] {};
  \node[right=0.1cm of 22] {\((T_2, 1):\) write y};

  \path[every node/.style={font=\sffamily\small,
        fill=white,inner sep=1pt}]
    (21) edge (22);

  \path[every node/.style={font=\sffamily\small,
        fill=white,inner sep=1pt}]
    (12) edge node [above=0.3cm] {[x==y]} (21)
    (13) edge node [above=0.3cm] {[x==y]} (22);

\end{tikzpicture}

  \caption{Example symbolic trace for the program of
    Figure~\ref{fig:example-program}. In all executions that adhere to
    this trace, the assertion in line 6 of thread \(T_1\)
    is not violated. (Transitive edges are omitted.)}
  \label{fig:example-trace}
\end{figure}
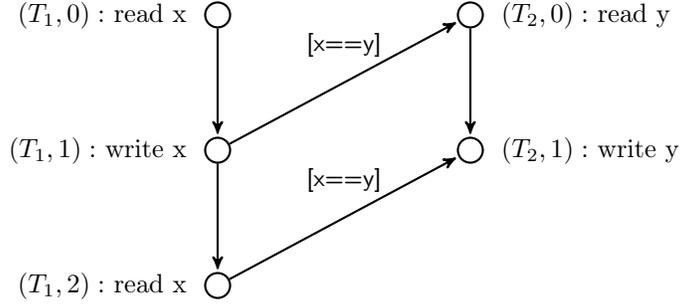

\begin{example}
  A symbolic trace for the program of
  Figure~\ref{fig:example-program}, is given in
  Figure~\ref{fig:example-trace}. The program consists of two threads,
  \(T_1\)
  and \(T_2\).
  Each thread is given a pointer as input and increments the value at
  the pointer's target, mistakenly without synchronization. Thread
  \(T_1\)
  asserts that the target of \lstinline{x} indeed holds the intended
  value. In case the pointers \lstinline{x} and \lstinline{y} point to
  different memory locations, the threads do not interfere with each
  other and the assertion holds. Otherwise, dependent accesses occur
  and the assertion does not hold under every possible ordering of
  events. The symbolic trace in Figure~\ref{fig:example-trace} ensures
  that the assertion holds in all executions that adhere to the
  trace. Nodes correspond to events and are labeled with the
  corresponding memory access for clarity. Edges between events of the
  same thread represent the thread's program order; edges between
  events of different threads represent scheduling constraints. In any
  case, all events are executed. Since dependencies between \(T_1\)
  and \(T_2\)
  occur only if the pointer targets match, the scheduling constraints
  are labeled with \lstinline{x==y}. In case \lstinline{x} \(\neq\)
  \lstinline{y}, the scheduling constraints do not apply and the
  ordering between threads is arbitrary. Otherwise, \((T_1, 1)\)
  has to be executed before \((T_2, 0)\)
  and \((T_1, 2)\) has to be executed before \((T_2, 1)\).
\end{example}

If a trace graph is used to encode scheduling constraints,
\((\sinit{0}, \xu/) \traceprefix/ \xo/\)
can be checked as follows: \((\sinit{0}, \xu/) \traceprefix/ \xo/\)
holds if \xu/ is empty. Intuitively, that \xu/ is empty means that it
does not contain any events that can violate any constraint given by
\xo/. If \xu/ is not empty, \((\sinit{0}, \xu/) \traceprefix/ \xo/\)
is satisfied if the first event \(\xe/_i\)
of \xu/ occurs in \xo/ without incoming edge that satisfies the
current path constraints and recursively,
\((\xs/', \xu/') \traceprefix/ \xo/'\)
holds, where \(\xs/'\)
is the successor state of \sinit{0} after \(\xe/_i\),
\(\xu/'\)
is \(\xu/\)
with the first event removed, and \(\xo/'\)
is \(\xo/\) with \(\xe/_i\) and all adjacent edges removed.

\subsection{Algorithm}\label{sec:algorithm}
The general IRS algorithm~\cite{Metzler2017} requires a
synchronization between individual threads and the IRS execution
environment after each event in order to check compliance of the
current execution with a previously verified trace. Additionally, it
stores all current admissible traces explicitly, which increases space
requirements and look-up times as the verification advances. With
Algorithm~\ref{alg:optimized-irs}, we present an IRS algorithm that
can be efficiently implemented. It addresses both previously described
issues by the use of trace prefixes as scheduling constraints and
allowing threads to run uninterrupted for multiple memory events
whenever scheduling constraints do not require synchronization.

In order to simplify the presentation, it is assumed that
the IRS execution environment enforces sequential consistency
independently from scheduling constraints. For platforms where this
incurs a considerable slowdown, scheduling of events of the same
thread can be relaxed by considering intra-thread scheduling
constraints.

\begin{algorithm}[t]
  \caption{IRS with trace prefixes and execution of sequences without
    synchronization}
\label{alg:optimized-irs}
\KwData{\admissibleprefix/ -- the admissible trace prefix, initially
  an arbitrary complete, correct trace of program \xP/}
\Verifier{
  initialize internal verifier state \xG/ such that \safe{\admissibleprefix/}\;
  \While{not \finished/}{
    do next verification step and update \xG/\;
    \If{\(\exists \xo/' < \admissibleprefix/ \ldotp \safe{\xo/'}\) \label{line:safe}}{
      \(\admissibleprefix/ \leftarrow \xo/'\)\;
    }
  }
}

\ExecutionEnvironment{
  set the current execution (\sinit{0}, \xu/) to the current program
  input and empty sequence\;
  \While{\xP/ has not terminated}{
    choose some sequence \xv/ from \free{\sinit{0}, \xu/,
    \admissibleprefix/}\;\label{line:optimized-irs:check}
    execute \xv/\;
    append \xv/ to \xu/\;
  }
}
\end{algorithm}

\textbf{Do Not Synchronize Already Reversed Races.} By using trace
prefixes as scheduling constraints, it is possible to avoid
synchronization before events when every possible continuation of the
current execution is proven to be error-free. The corresponding part
in an admissible trace does not have to be enforced and scheduling
constraints can be removed.

Algorithm~\ref{alg:basic} maintains a set \xV/ of admissible
traces. The verifier may add any safe trace to \xV/. Instead of
managing a set of admissible traces, Algorithm~\ref{alg:optimized-irs}
uses a single trace \admissibleprefix/ as the current admissible trace
prefix. Every event that occurs in this prefix has to be executed
according to its partial order, however every additional event may be
executed without synchronization. Once the verifier has collected
enough information about correct executions of the program, the
admissible trace prefix is updated. In order to prevent unnecessary
assumptions on the verifier, we do not require the use of a specific
data structure such as a state graph. Instead, we only require that
the verifier maintains an internal state \xG/ that contains
information on safe parts of the state space.

\begin{definition}
  Given a program \xP/ and a state predicate \errorfree{} (induced by
  the property to be verified), a \emph{verifier} maintains an
  internal state \xG/ and provides predicates \safe{} and \finished/
  such that
  \(\forall \xs/ \ldotp \safe{\xs/} \Rightarrow \forall \xu/ \ldotp
  \forall \xs/' \ldotp \xs/ \transition{\xu/}{} \xs/' \Rightarrow
  \errorfree{\xs/'}\)
  and \finished/ holds when verification has finished.
\end{definition}

In other words, a correct verifier guarantees safety for all (partial)
executions from a state \xs/ whenever \safe{\xs/} holds. We use a
derived definition for safety of trace prefixes \xo/ that guarantees
safety for all executions satisfying \xo/: \safe{\xo/} holds if
\(\forall \sinit{0} \in \initstates/ \ldotp \forall \xs/' \ldotp
\forall \xu/ \ldotp \sinit{0} \transition{\xu/}{} \xs/' \Rightarrow
(((\sinit{0}, \xu/) \tracematch/ \xo/ \Rightarrow \safe{\xs/'}) \wedge
((\sinit{0}, \xu/) \traceprefix/ \xo/ \Rightarrow
\errorfree{\xs/'}))\).
An implementation of the verifier may, for example, use an abstract
reachability tree~\cite{Wachter2013} to realize~\xG/.

The current admissible trace prefix \admissibleprefix/ is updated by
shortening it, i.e., by removing constraints at the end of its
happens-before order. Formally, a new admissible trace prefix
\(\xo/'\)
is required to satisfy \(\xo/' < \admissibleprefix/\),
which is defined as:
\((\events{1}, \transitions{1}) < (\events{2}, \transitions{2})\)
if
\(\events{1} \subsetneq \events{2} \wedge \transitions{1} = \{(\xe/_1,
\xc/, \xe/_2) \in \transitions{2} : \xe/_1, \xe/_2 \in \events{1}\}
\wedge \forall \xe/_1 \transitions{2} \xe/_2 \ldotp \xe/_1 \notin
\events{1} \Rightarrow \xe/_2 \notin \events{1}\).
On a more abstract level, the verifier finds an initial, complete, and
correct trace \xo/ of the program and generates a sequence
\(\xo/ > \xo/_1 > \dotsc > \xo/_n\)
of subsequent trace prefixes such that for all \(1 \leq i \leq n\),
\safe{\xo/_i}.

A verifier can update a trace prefix \xo/ as follows. Each edge
\((\xe/_1, \xe/_2)\)
with \(\tid{\xe/_1} \neq \tid{\xe/_2}\)
in \xo/ is interpreted as a scheduling constraint that requires
\(\xe/_2\)
to be executed after \(\xe/_1\).
Updates of trace prefixes remove scheduling constraints. Let \(\xo/'\)
be \(\xo/\)
with \(\xe/_1\),
\(\xe/_2\),
and all their successors (w.r.t. the happens-before relation)
removed. It is safe to remove the scheduling constraint
\((\xe/_1, \xe/_2)\)
if all states \xs/ that are reachable by a linearization of \(\xo/'\)
are safe, i.e., \safe{\xs/}. Depending on the verification approach
used, it may be more efficient to delay the removal of
\((\xe/_1, \xe/_2)\)
until it occurs at an end of \xo/, w.r.t. the happens-before relation,
i.e., no event happens after \(\xe/_2\)
that has an incoming or outgoing edge with an event from an other
thread.

In the worst case, even if scheduling constraint \((\xe/_1, \xe/_2)\)
is at the end of a trace prefix, the verifier has to prove safety for
exponentially many states before \((\xe/_1, \xe/_2)\)
can be safely removed. On the one hand, this complexity is a general
limitation of IRS. On the other hand, the duty of the verifier can be
reduced exponentially by adding only one scheduling constraint, which
may reduce the verification delay considerably.

\textbf{Do Not Preempt Minimal Events.} In addition to the use of
trace prefixes, Algorithm~\ref{alg:optimized-irs} omits
synchronization before events that do not have to occur second in a
race, i.e., events that do not have a predecessor in
\admissibleprefix/ from a different thread, i.e., events
\(\{ \xe/ \in \xo/ : \forall \xe/' \in \xo/ \ldotp \xe/'
\transition{\xc/}{\xo/} \xe/ \Rightarrow \tid{\xe/'} = \tid{\xe/}\}\).

The execution environment of Algorithm~\ref{alg:optimized-irs} reduces
the number of synchronizations by permitting a sequence of events,
potentially from multiple threads, between two synchronizations. This
sequence is chosen from the set \free{\sinit{0}, \xu/, \xo/} as a
continuation of the current execution \((\sinit{0}, \xu/)\)
that adheres to \xo/ or contains only synchronization-free events. The
definition of \free{} reflects both optimizations, trace prefixes and
execution of sequences between two synchronizations.

\begin{definition}
  Given an execution \((\sinit{0}, \xu/)\)
  with \(\xu/ = \xe/_1 \dotsc \xe/_m\),
  the set of synchronization-free events, \free{\sinit{0}, \xu/,
    \xo/}, is defined as
  \(\free{\sinit{0}, \xu/, \xo/} := \{ \xe/_{m+1} \dotsc \xe/_{n} \in
  (\alphabet{} \times \xnaturals/)^+ : \exists \xs/ \ldotp \sinit{0}
  \transition{\xu/ \concatenate/ \xe/_{m+1} \dotsc \xe/_{n}}{} \xs/
  \wedge \forall m < i \leq n \ldotp (\xe/_i \notin \xo/ \vee
  (\sinit{0}, \xu/ \concatenate/ \xe/_1 \dotsc \xe/_i) \traceprefix/
  \xo/)\).
\end{definition}

In Algorithm~\ref{alg:optimized-irs}, \free{} replaces the check for a
suitable admissible trace as in Algorithm~\ref{alg:basic}. In
line~\ref{line:optimized-irs:check}, Algorithm~\ref{alg:optimized-irs}
selects a sequence \xv/ from \free{\sinit{0}, \xu/,
  \admissibleprefix/} and executes all events in \xv/ without
intermediate synchronization.

\subsection{Correctness and Progress}
An IRS algorithm is correct if only safe executions, i.e., executions
that do not violate the safety specification, can occur under its
execution environment. The following theorem provides correctness of
Algorithm~\ref{alg:optimized-irs}.
\begin{theorem}[Correctness]
  Whenever an execution \((\sinit{0}, \xu/)\)
  with \(\xu/ = \xe/_1 \dotsc \xe/_n\)
  and
  \(\sinit{0} \transition{\xe/_1}{} \dotsc \transition{\xe/_n}{}
  \xs/_n\)
  has been executed by Algorithm~\ref{alg:optimized-irs}, all visited
  states are error-free, i.e.,
  \(\forall 0 \leq i \leq n \ldotp \errorfree{\xs/_i}\).
  \begin{proof}
    Let \((\sinit{0}, \xu/)\)
    with \(\xu/ = \xe/_1 \dotsc \xe/_n\)
    and
    \(\sinit{0} \transition{\xe/_1}{} \dotsc \transition{\xe/_n}{}
    \xs/_n\)
    be an execution executed by Algorithm~\ref{alg:optimized-irs}, let
    \admissibleprefix/ be the final admissible trace prefix, and \xG/
    the final verifier state. By the definition of \free{}, there
    exists a prefix \(\xv/ = \xe/_1 \dotsc \xe/_k\) of
    \xu/ such that
    \((\sinit{0}, \xv/) \tracematch/ \admissibleprefix/\).
    Algorithm~\ref{alg:optimized-irs} in line~\ref{line:safe} ensures
    that \safe{\admissibleprefix/}. By the definition of \safe{},
    \safe{\xs/_k} and \errorfree{\xs/_i} holds for all
    \(0 \leq i \leq k\).
    The verifier guarantees that \safe{\xs/_k} implies
    \errorfree{\xs/_i} for all \(k \leq i \leq n\).
  \end{proof}
\end{theorem}

In addition to correctness, an important requirement is that a program
is never completely blocked by scheduling constraints (provided that
at least one correct execution exists). The following progress theorem
guarantees that this cannot happen with
Algorithm~\ref{alg:optimized-irs}.
\begin{theorem}[Progress]
  Whenever an execution \((\sinit{0}, \xu/)\)
  has been executed by Algorithm~\ref{alg:optimized-irs} with verified
  state graph \xG/ and admissible trace prefix \admissibleprefix/,
  either the program has terminated or \free{\sinit{0}, \xu/,
    \admissibleprefix/} is not empty.
\begin{proof}
  Assume that the program has not terminated.  Let \admissibleprefix/
  be the initial admissible trace prefix and let
  \(\admissibleprefix/'\)
  be the current admissible trace prefix. Case
  \((\sinit{0}, \xu/) \traceprefix/ \admissibleprefix/'\)
  and not \((\sinit{0}, \xu/) \tracematch/ \admissibleprefix/'\):
  as \admissibleprefix/ is a feasible trace and
  \(\admissibleprefix/' < \admissibleprefix/\)
  (no constraints can be added), there exists some
  \(\xe/ \in \admissibleprefix/'\)
  such that
  \((\sinit{0}, \xu/ \concatenate/ \xe/) \traceprefix/
  \admissibleprefix/\).
  By definition,
  \(\xe/ \in \free{\sinit{0}, \xu/, \admissibleprefix/'}\).
  Case not \((\sinit{0}, \xu/) \traceprefix/ \admissibleprefix/'\):
  by correctness,
  \((\sinit{0}, \xv/) \traceprefix/ \admissibleprefix/'\)
  for some prefix \xv/ of \xu/. Hence, for any \xe/ such that
  \((\sinit{0}, \xv/ \concatenate/)\)
  is an execution,
  \(\xe/ \in \free{\sinit{0}, \xu/, \admissibleprefix/'}\).
\end{proof}
\end{theorem}



\section{Implementation}\label{sec:implementation}
We have implemented Algorithm~\ref{alg:optimized-irs} from
Section~\ref{sec:optimization} in our IRS
prototype~\cite{Metzler2017}. This prototype handles C and C++
programs translated to LLVM-IR. The LLVM-IR code is instrumented via
the LLVM compiler infrastructure~\cite{LLVM} in order to enforce an
admissible trace prefix whenever the program is executed. The IRS
execution environment is realized completely inside the instrumented
application program and does not depend on any modifications of the
operating system or assumptions on the used scheduler. Via a standard
dependency analysis the prototype identifies all \emph{dependent}
memory accesses, which are memory accesses that either directly access
global memory or may influence the result of an other global memory
access. Scheduling constraints are enforced by callbacks directly
before each dependent memory access that check whether this memory
access is currently permitted. Callbacks directly after each dependent
memory access communicate to other threads that the memory access has
been performed. Memory fences inside these callbacks ensure sequential
consistency, as assumed by our presentation in
Section~\ref{sec:algorithm}. Before each instrumented memory access, a
thread checks whether an event of an other thread has to occur before
its own upcoming event via a look-up in a global vector clock. Busy
waiting is performed until the current thread is permitted to
continue. After the memory access, the callback signals that the
memory access is completed by updating the global vector clock. In
contrast to earlier versions of our prototype, no thread is added to
the program.

When testing several alternatives of implementing schedule
enforcement, we observed that, as expected, lock-based implementations
of waiting for other threads' events are much slower than busy
waiting. A disadvantage of busy waiting is CPU consumption during
waiting, which can reduce performance when more threads are active
than hardware cores are available. We expect that improvements over
our current, simple scheme of busy-waiting for permissions can be made
by the use of a more advanced combination of busy waiting with
lock-based synchronization or scheduler interaction (e.g., the POSIX
\lstinline{sched_yield()} system call). Additionally, we tested an
implementation that uses a loadable kernel module to communicate with
the Linux scheduler. Whenever an event is not yet permitted to be
executed, the corresponding task's state is set to
\lstinline{TASK_WAIT} and only restored once the event is
permitted. This design circumvents the additional CPU consumption of
busy waiting. However, additional overhead appears because the current
program counter of each thread has to be communicated to the loadable
kernel module. In our tests, this design showed only an advantage if
most events were constrained, i.e., the likelihood that an event has
to wait is high.


\section{Experimental Evaluation}\label{sec:evaluation}
\begin{figure}[t]
  \centering
  \begin{tikzpicture}[gnuplot]
\tikzset{every node/.append style={font={\small}}}
\path (0.000,0.000) rectangle (12.500,8.750);
\gpcolor{color=gp lt color border}
\gpsetlinetype{gp lt border}
\gpsetdashtype{gp dt solid}
\gpsetlinewidth{1.00}
\draw[gp path] (1.504,0.985)--(1.684,0.985);
\draw[gp path] (11.947,0.985)--(11.767,0.985);
\node[gp node right] at (1.320,0.985) {$0.95$};
\draw[gp path] (1.504,2.228)--(1.684,2.228);
\draw[gp path] (11.947,2.228)--(11.767,2.228);
\node[gp node right] at (1.320,2.228) {$1$};
\draw[gp path] (1.504,3.470)--(1.684,3.470);
\draw[gp path] (11.947,3.470)--(11.767,3.470);
\node[gp node right] at (1.320,3.470) {$1.05$};
\draw[gp path] (1.504,4.713)--(1.684,4.713);
\draw[gp path] (11.947,4.713)--(11.767,4.713);
\node[gp node right] at (1.320,4.713) {$1.1$};
\draw[gp path] (1.504,5.956)--(1.684,5.956);
\draw[gp path] (11.947,5.956)--(11.767,5.956);
\node[gp node right] at (1.320,5.956) {$1.15$};
\draw[gp path] (1.504,7.198)--(1.684,7.198);
\draw[gp path] (11.947,7.198)--(11.767,7.198);
\node[gp node right] at (1.320,7.198) {$1.2$};
\draw[gp path] (1.504,8.441)--(1.684,8.441);
\draw[gp path] (11.947,8.441)--(11.767,8.441);
\node[gp node right] at (1.320,8.441) {$1.25$};
\draw[gp path] (11.947,0.985)--(11.947,1.165);
\draw[gp path] (11.947,8.441)--(11.947,8.261);
\node[gp node center] at (11.947,0.677) {$0\%$};
\draw[gp path] (9.858,0.985)--(9.858,1.165);
\draw[gp path] (9.858,8.441)--(9.858,8.261);
\node[gp node center] at (9.858,0.677) {$20\%$};
\draw[gp path] (7.770,0.985)--(7.770,1.165);
\draw[gp path] (7.770,8.441)--(7.770,8.261);
\node[gp node center] at (7.770,0.677) {$40\%$};
\draw[gp path] (5.681,0.985)--(5.681,1.165);
\draw[gp path] (5.681,8.441)--(5.681,8.261);
\node[gp node center] at (5.681,0.677) {$60\%$};
\draw[gp path] (3.593,0.985)--(3.593,1.165);
\draw[gp path] (3.593,8.441)--(3.593,8.261);
\node[gp node center] at (3.593,0.677) {$80\%$};
\draw[gp path] (1.504,0.985)--(1.504,1.165);
\draw[gp path] (1.504,8.441)--(1.504,8.261);
\node[gp node center] at (1.504,0.677) {$100\%$};
\draw[gp path] (1.504,8.441)--(1.504,0.985)--(11.947,0.985)--(11.947,8.441)--cycle;
\node[gp node center,rotate=-270] at (0.276,4.713) {Relative execution time};
\node[gp node center] at (6.725,0.215) {Scheduling constraints};
\node[gp node right] at (10.479,8.107) {Bigshot};
\gpcolor{rgb color={0.580,0.000,0.827}}
\draw[gp path] (10.663,8.107)--(11.579,8.107);
\draw[gp path] (1.504,3.470)--(11.947,2.973);
\gpsetpointsize{4.80}
\gppoint{gp mark 1}{(1.504,3.470)}
\gppoint{gp mark 1}{(11.947,2.973)}
\gppoint{gp mark 1}{(11.121,8.107)}
\gpcolor{color=gp lt color border}
\node[gp node right] at (10.479,7.799) {Dekker};
\gpcolor{rgb color={0.000,0.620,0.451}}
\draw[gp path] (10.663,7.799)--(11.579,7.799);
\draw[gp path] (1.504,3.222)--(6.726,2.973)--(11.947,2.725);
\gppoint{gp mark 2}{(1.504,3.222)}
\gppoint{gp mark 2}{(6.726,2.973)}
\gppoint{gp mark 2}{(11.947,2.725)}
\gppoint{gp mark 2}{(11.121,7.799)}
\gpcolor{color=gp lt color border}
\node[gp node right] at (10.479,7.491) {Fibonacci};
\gpcolor{rgb color={0.337,0.706,0.914}}
\draw[gp path] (10.663,7.491)--(11.579,7.491);
\draw[gp path] (1.504,5.459)--(7.248,4.464)--(9.441,5.210)--(11.947,3.719);
\gppoint{gp mark 3}{(1.504,5.459)}
\gppoint{gp mark 3}{(7.248,4.464)}
\gppoint{gp mark 3}{(9.441,5.210)}
\gppoint{gp mark 3}{(11.947,3.719)}
\gppoint{gp mark 3}{(11.121,7.491)}
\gpcolor{color=gp lt color border}
\node[gp node right] at (10.479,7.183) {Lamport};
\gpcolor{rgb color={0.902,0.624,0.000}}
\draw[gp path] (10.663,7.183)--(11.579,7.183);
\draw[gp path] (1.504,5.210)--(2.131,5.210)--(5.368,5.459)--(7.352,5.459)--(7.979,5.459)%
  --(9.336,5.210)--(10.589,5.210)--(11.320,5.459)--(11.947,2.973);
\gppoint{gp mark 4}{(1.504,5.210)}
\gppoint{gp mark 4}{(2.131,5.210)}
\gppoint{gp mark 4}{(5.368,5.459)}
\gppoint{gp mark 4}{(7.352,5.459)}
\gppoint{gp mark 4}{(7.979,5.459)}
\gppoint{gp mark 4}{(9.336,5.210)}
\gppoint{gp mark 4}{(10.589,5.210)}
\gppoint{gp mark 4}{(11.320,5.459)}
\gppoint{gp mark 4}{(11.947,2.973)}
\gppoint{gp mark 4}{(11.121,7.183)}
\gpcolor{color=gp lt color border}
\node[gp node right] at (10.479,6.875) {Peterson};
\gpcolor{rgb color={0.941,0.894,0.259}}
\draw[gp path] (10.663,6.875)--(11.579,6.875);
\draw[gp path] (1.504,4.216)--(2.966,4.464)--(3.697,3.719)--(11.529,3.967)--(11.947,1.731);
\gppoint{gp mark 5}{(1.504,4.216)}
\gppoint{gp mark 5}{(2.966,4.464)}
\gppoint{gp mark 5}{(3.697,3.719)}
\gppoint{gp mark 5}{(11.529,3.967)}
\gppoint{gp mark 5}{(11.947,1.731)}
\gppoint{gp mark 5}{(11.121,6.875)}
\gpcolor{color=gp lt color border}
\node[gp node right] at (10.479,6.567) {Shared Pointer};
\gpcolor{rgb color={0.000,0.447,0.698}}
\draw[gp path] (10.663,6.567)--(11.579,6.567);
\draw[gp path] (1.504,7.695)--(4.950,7.447)--(8.501,7.198)--(11.947,3.222);
\gppoint{gp mark 6}{(1.504,7.695)}
\gppoint{gp mark 6}{(4.950,7.447)}
\gppoint{gp mark 6}{(8.501,7.198)}
\gppoint{gp mark 6}{(11.947,3.222)}
\gppoint{gp mark 6}{(11.121,6.567)}
\gpcolor{color=gp lt color border}
\node[gp node right] at (10.479,6.259) {uninstrumented};
\gpcolor{rgb color={0.753,0.753,0.753}}
\draw[gp path] (10.663,6.259)--(11.579,6.259);
\draw[gp path] (1.504,2.228)--(1.609,2.228)--(1.715,2.228)--(1.820,2.228)--(1.926,2.228)%
  --(2.031,2.228)--(2.137,2.228)--(2.242,2.228)--(2.348,2.228)--(2.453,2.228)--(2.559,2.228)%
  --(2.664,2.228)--(2.770,2.228)--(2.875,2.228)--(2.981,2.228)--(3.086,2.228)--(3.192,2.228)%
  --(3.297,2.228)--(3.403,2.228)--(3.508,2.228)--(3.614,2.228)--(3.719,2.228)--(3.825,2.228)%
  --(3.930,2.228)--(4.036,2.228)--(4.141,2.228)--(4.247,2.228)--(4.352,2.228)--(4.458,2.228)%
  --(4.563,2.228)--(4.669,2.228)--(4.774,2.228)--(4.880,2.228)--(4.985,2.228)--(5.090,2.228)%
  --(5.196,2.228)--(5.301,2.228)--(5.407,2.228)--(5.512,2.228)--(5.618,2.228)--(5.723,2.228)%
  --(5.829,2.228)--(5.934,2.228)--(6.040,2.228)--(6.145,2.228)--(6.251,2.228)--(6.356,2.228)%
  --(6.462,2.228)--(6.567,2.228)--(6.673,2.228)--(6.778,2.228)--(6.884,2.228)--(6.989,2.228)%
  --(7.095,2.228)--(7.200,2.228)--(7.306,2.228)--(7.411,2.228)--(7.517,2.228)--(7.622,2.228)%
  --(7.728,2.228)--(7.833,2.228)--(7.939,2.228)--(8.044,2.228)--(8.150,2.228)--(8.255,2.228)%
  --(8.361,2.228)--(8.466,2.228)--(8.571,2.228)--(8.677,2.228)--(8.782,2.228)--(8.888,2.228)%
  --(8.993,2.228)--(9.099,2.228)--(9.204,2.228)--(9.310,2.228)--(9.415,2.228)--(9.521,2.228)%
  --(9.626,2.228)--(9.732,2.228)--(9.837,2.228)--(9.943,2.228)--(10.048,2.228)--(10.154,2.228)%
  --(10.259,2.228)--(10.365,2.228)--(10.470,2.228)--(10.576,2.228)--(10.681,2.228)--(10.787,2.228)%
  --(10.892,2.228)--(10.998,2.228)--(11.103,2.228)--(11.209,2.228)--(11.314,2.228)--(11.420,2.228)%
  --(11.525,2.228)--(11.631,2.228)--(11.736,2.228)--(11.842,2.228)--(11.947,2.228);
\gpcolor{color=gp lt color border}
\draw[gp path] (1.504,8.441)--(1.504,0.985)--(11.947,0.985)--(11.947,8.441)--cycle;
\gpdefrectangularnode{gp plot 1}{\pgfpoint{1.504cm}{0.985cm}}{\pgfpoint{11.947cm}{8.441cm}}
\end{tikzpicture}
  \caption{Execution time overhead of IRS relative to uninstrumented
    benchmarks for decreasing numbers of scheduling constraints
    (two-threaded benchmarks)}
  \label{fig:results-twothreaded}
\end{figure}
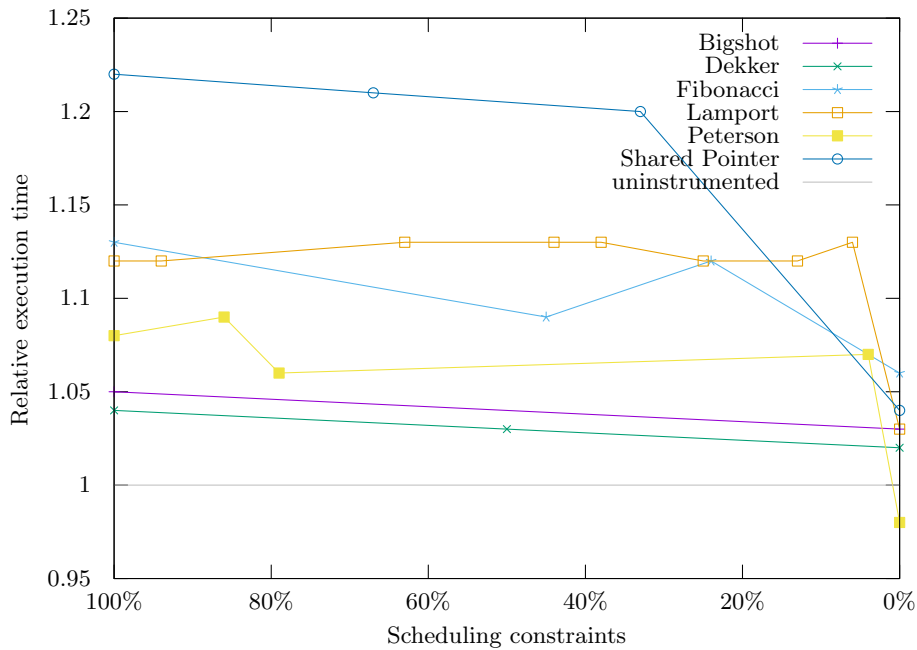

Enforcing scheduling constraints in order to disable schedules outside
of a given admissible trace is likely to incur execution time overhead
(here: simply \emph{overhead}) in comparison to plain program
executions (without IRS). A crucial factor for the applicability of
IRS in practice is how scheduling constraints in IRS influence this
overhead, which we evaluate on several benchmark programs. The main
goal of this evaluation is to investigate whether, for a given
admissible trace and induced scheduling constraints, relaxing those
constraints reduces the overhead and, if this is the case, how
fast. Additionally, we investigate whether the selection of the
initial and following admissible traces, i.e., the structure of the
admissible trace prefix, influences the overhead. Our implementation
of IRS and the benchmarks we use are publicly
available\footnote{\url{https://www.informatik.tu-darmstadt.de/media/deeds/pdf_patrick/irs-bin.7z.zip}}.

\textbf{Setup.} All experiments have been conducted with our IRS
implementation described in Section~\ref{sec:implementation}. The
hardware used is an Intel Core~i5-6500 CPU at~3.20GHz with four cores
running Linux~4.8.0. Each benchmark is run with and without
instrumentation by our prototype. The instrumented version is run in
several configurations, with a decreasing amount of scheduling
constraints. The initial number of scheduling constraints and the
number of scheduling constraints that can be removed in one step, and
thereby the number of configurations per benchmark, vary as the number
of conflicting memory accesses varies among benchmarks. Each
configuration is run 1000 times. We report the median execution time
and overhead relative to the unmodified benchmark. Detailed
measurement results are shown in Appendix~\ref{app:results}.

\textbf{Benchmark set 1.} The first set of benchmarks are concurrent
programs from the SV-COMP benchmark suite~\cite{SV-COMP} (Bigshot,
Dekker, Fibonacci, Lamport, and Peterson) and the POR literature
(Shared Pointer, ~\cite{Gueta2007}). These benchmarks contain standard
algorithms for concurrency problems, typical concurrency bugs, and
patterns with many concurrent accesses. Their size lies between 74 and
130 lines of code. We chose these benchmarks because they are
well-studied verification problems and contain a high amount of
concurrent interaction relative to their size. In contrast to larger
benchmarks with fewer concurrent memory accesses in relation to the
total number of statements, these benchmarks are suitable to highlight
performance issues of IRS: only concurrent accesses may be slowed down
by IRS; more statements without concurrent accesses are expected to
reduce the performance impact of IRS. All benchmarks contain two
threads.

The results for Benchmark set~1 are shown in
Fig.~\ref{fig:results-twothreaded}. For these benchmarks, IRS produces
a maximum overhead of 22\%, which is much less than we expected and
might be already an acceptable overhead for certain applications. For
all benchmarks, the overhead is reduced by relaxing scheduling
constraints, albeit in some cases, a significant reduction occurs only
at the last reduction step. In some cases, the overhead is negative,
i.e., the instrumented version of a benchmark executed faster than the
plain benchmark. We conjecture that both measurement noise as well as
improved timing of cache operations due to a different interleaving of
memory operations may be relevant for this effect, as already noted by
Olszewski et al.~\cite{Olszewski2009}. Similarly, an increased
overhead after removing scheduling constraints could be caused in such
a way. Overall, both the initial overhead and the amount of reductions
are lower than we expected.

\begin{figure}[t]
  \centering
  \input{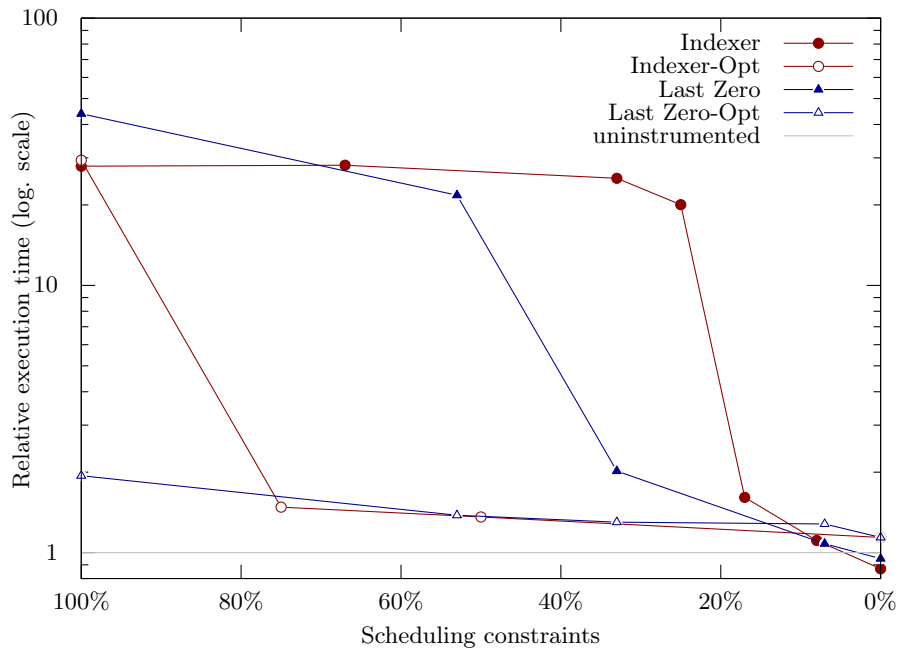}
  \caption{Execution time overhead of IRS relative to uninstrumented
    benchmarks for decreasing numbers of scheduling constraints
    (many-threaded benchmarks)}
  \label{fig:results-manythreaded}
\end{figure}

\textbf{Benchmark set 2.} Since we expected a higher overhead, we
conduct the same experiment on two benchmarks from the POR literature:
Indexer~\cite{Flanagan2005} with 15~threads, a benchmark simulating
accesses to a shared hash map, and Last Zero~\cite{Abdulla2014} with
16~threads, where the first thread searches the last zero in an
integer array written by all other threads. For these two benchmarks,
we expect a higher overhead as a larger amount of threads and
dependencies results in a higher amount of scheduling
constraints. Fig.~\ref{fig:results-manythreaded} shows the
corresponding results. Indeed, for the Indexer and Last Zero
benchmarks, the overhead is much higher. Interestingly, the overhead
for Indexer abruptly decreases from 1904\% to 61\% at the transition
from three to two scheduling constraints. We explain this observation
by the fact that the permitted trace prefix with three scheduling
constraints requires three threads to wait, while after removing one
scheduling constraint, only two threads have to wait. Since our
implementation uses busy waiting, many concurrently waiting threads
may prevent threads that are not required to wait from quickly
proceeding.

\begin{figure*}[t]
  \centering
  \subfloat[Many threads wait rarely]{
    \label{fig:trace-horizontal}
    \begin{minipage}{0.45\textwidth}
      \centering
      \includegraphics[scale=0.125]{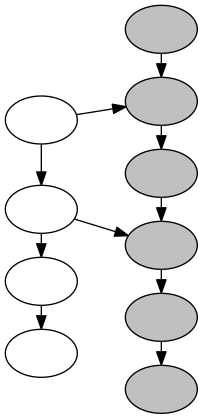}
      \includegraphics[scale=0.125]{figures/trace-indexer-t15-horizontal-0016.png}
      \includegraphics[scale=0.125]{figures/trace-indexer-t15-horizontal-0016.png}
      \includegraphics[scale=0.125]{figures/trace-indexer-t15-horizontal-0016.png}
    \end{minipage}}
  \subfloat[Few threads wait often (optimized)]{
    \label{fig:trace-vertical}
    \begin{minipage}{0.45\textwidth}
      \centering
      \includegraphics[scale=0.125]{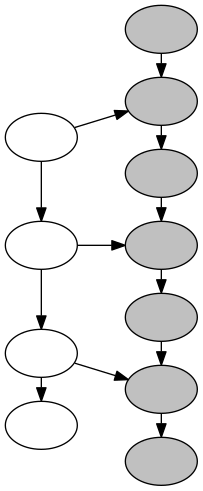}
      \includegraphics[scale=0.125]{figures/trace-indexer-t15-vertical-0008.png}
      \includegraphics[scale=0.125]{figures/trace-indexer-t15-vertical-0008.png}
    \end{minipage}}
  \caption{\label{fig:traces}Trace prefixes for Indexer (threads with
    only conflict-free events are omitted)}
\end{figure*}

\textbf{Structure of scheduling constraints.} An interesting question
is whether the overhead can be reduced by choosing a different trace
prefix with roughly the same amount of scheduling
constraints. Interestingly, we have found optimized traces for both
Indexer and Last Zero that indeed show a drastically reduced overhead
with the same or even more scheduling constraints. The corresponding
results are depicted as Indexer-Opt and Last Zero-Opt in
Fig.~\ref{fig:results-manythreaded}. For Indexer, we found that
choosing a trace prefix that requires less threads to wait can be
executed faster. Fig.~\ref{fig:traces} shows two alternative traces
for Indexer. Nodes represent events and edges a happens-before
relation. The nodes of even-indexed threads are shown in gray and
events of the same thread are arranged vertically one below the other.
Fig.~\ref{fig:trace-horizontal} shows one of the slower trace
prefixes, where many threads wait rarely, and
Fig.~\ref{fig:trace-vertical} shows one of the faster (optimized)
trace prefixes, where few threads wait often. For the former trace
prefix, 16~Mazurkiewicz traces, for the latter trace prefix, only
eight~Mazurkiewicz traces have to be verified. Although more
scheduling constraints are enforced, the program execution is faster
with the latter trace prefix. While we optimize trace prefixes
manually, it is conceivable that verifiers can prioritize faster trace
prefixes automatically, e.g., by applying a heuristic or by testing
few traces and comparing their overhead. Such a prioritization
resembles the effects of ordering heuristics on the performance of POR
algorithms studied by Lauterburg et al.~\cite{Lauterburg2010}.  For
Last Zero, our original trace prefixes require the second event of a
worker thread to wait for the first event of the next worker
thread. By letting threads wait already before their first events, the
program execution is drastically accelerated already for 100\%
scheduling constraints, i.e., when only a single Mazurkiewicz trace is
verified.

\textbf{Summary.} Our results show that relaxing scheduling
constraints can reduce the overhead for all benchmarks. For example,
after verifying only~eight of~4096 Mazurkiewicz traces of Indexer, the
overhead is reduced from~2841\% to~48\%. However, in other cases, the
execution time may not decrease considerably until a large part of all
scheduling constraints have been removed. In yet other cases, the
overhead is reduced considerably by removing a single scheduling
constraint, while it does not change considerably before and after
this step. Besides the number of scheduling constraints, the choice of
the permissible trace prefix, i.e., the structure of the induced
scheduling constraints, may have a large influence on the
overhead. These observations suggest that a sensible selection of an
initial trace during verification can considerably improve the
execution time performance of a program that is executed with IRS.
Comparing our current results for Indexer and Last Zero to earlier
experiments with a less optimized schedule
enforcement~\cite{Metzler2017}, we see a considerable speed-up when
optimized trace prefixes are used.

\section{Related Work}
Deterministic multi-threading (DMT) \cite{Bergan2010, Cui2013,
  Cui2011, Liu2011, Olszewski2009, Aviram2010} limits the amount of
non-determinism due to scheduling for multi-threaded programs.
Dthreads by Liu et al.~\cite{Liu2011} adapts the interface of the
multi-threading library Pthreads and guarantees, for any given input,
a deterministic execution. Dthreads interleaves parallel phases (in
which threads write only to a local copy of the shared memory) and
sequential phases (in which the local copies are merged). Dthreads
cannot handle programs that bypass the Pthreads library by
synchronizing directly over shared memory~\cite{Liu2011}.

Cui et al.\ propose Peregrine~\cite{Cui2011}, which initially records
a set of executions and enforces schedules of these initial executions
during subsequent executions where these schedules are
compatible. Schedules may be incompatible if an input is seen that
leads to a different schedule. In the subsequently presented Parrot
framework~\cite{Cui2013}, Cui et al.\ propose to combine DMT with a
model checker for bug-finding. Parts of a program that are manually
marked as performance-critical are executed non-deterministically and
model checked to increase the confidence about their correctness. Only
the remaining parts of the program are executed deterministically, so
that the overhead of additional synchronization is reduced.

In contrast to IRS, the above described DMT approaches do not provide
any guarantees about \emph{which} schedule is enforced, for a
particular input. Using these approaches to simplify program
verification is therefore impractical if many program inputs need to
be covered. While we conjecture that some of the former techniques can
be extended to communicate a general scheduling policy that guides a
verifier, it is not directly clear how to do so. In contrast, IRS
provides a formal interface that uses admissible traces to communicate
scheduling constraints. Additionally, the above described DMT
approaches do not allow to relax scheduling constraints during
runtime, in contrast to IRS, which enables to iteratively relax
scheduling constraints and, provided that the program is eventually
proven safe, remove all scheduling constraints. On the implementation
level, the approaches of~\cite{Olszewski2009,Liu2011,Cui2013} (but
not~\cite{Bergan2010,Cui2011}) synchronize only at library calls (such
as uses of Pthreads locks), which improves execution time performance
but may result in non-determinstic executions when global memory is
accessed (perhaps accidentally) directly, e.g., without lock
protection. In contrast, our IRS implementation schedules all accesses
to shared variables, which we consider to be important, as the task of
verification is to guaranty a safe executions without the assumption
that all memory accesses are protected by synchronization. The
approach of~\cite{Bergan2013} can handle inputs symbolically but again
cannot handle unprotected memory accesses.

Program analyses that use context bounding~\cite{Qadeer2005} consider
only those executions of a program which contain only up to \(k\)
context switches between threads, for a typically small bound \(k\).
While reachability for concurrent, recursive programs is
undecidable~\cite{Ramalingam2000}, additionally bounding the number of
context switches makes the problem
decidable~\cite{Qadeer2005}. Context bounding may be used within IRS,
although it is only a special case of scheduling constraints in
IRS. Similar to context bounding, a generally undecidable model
checking problem may become decidable when handled with IRS: by only
checking a limited set of symbolic traces, IRS enables to safely use a
program even if its reachability problem is undecidable under
unconstrained scheduling.

When applied with bounded model checking (BMC) for concurrency bug
finding~\cite{Rabinovitz2005, Cordeiro2011, Musuvathi2007, Lal2009,
  Inverso2014}, context bounding focuses the search for erroneous
schedules to those with few context switches. Consequently, potential
bugs are missed that manifest themselves only after more context
switches than the current bound. However, based on empirical results,
Musuvathi and Qadeer argue that a low context bound is sufficient to
find many interesting bugs~\cite{Musuvathi2007}. They propose
iterative context bounding (ICB) as an extension to BMC: a program is
iteratively checked with an increasing context bound, similar to
increasing the bound on execution lengths in BMC. Given limited
resources (that usually do not allow to search the complete state
space of a program), ICB prioritizes schedules with few context
switches. This search strategy of ICB could be used by a model checker
in conjunction with IRS. However, in contrast to bug finding based on
BMC, IRS requires a sound program analysis (under scheduling
constraints), i.e., a safety proof for complete, unbounded program
executions, which is not given, in general, by BMC. Another difference
between context bounding in bug finding and IRS are guarantees about
scheduling: when searching for erroneous schedules, bug finding may
use assumptions about the likelihood of schedules in order to guide
the search. However, any assumptions about the likelihood of schedules
are not enforced. Bug finding consequently accepts to miss feasible
executions of a program that contain, e.g., a bug that has not been
found under context bounding. In contrast, IRS guarantees that only
checked executions may occur.

Nguyen et al.\ \cite{Nguyen2017} transform a concurrent program into
several instances that show only a reduced number of schedules. Each
instance is checked individually by BMC (and with a context
bound). Similar to IRS, this decreases the complexity of the model
checking problem and improves bug finding. However, their approach of
dividing a program into instances is based on lazy sequentialization
for BMC~\cite{Fischer2013} and therefore not directly usable for
verification.

Conditional model checking~\cite{Beyer2012a} is a general framework to
reuse (in general arbitrary) intermediate verification results. In
contrast to IRS, it does not require intermediate verification results
to prove safety of a fully functional program variant and does not
enforce the preconditions of the intermediate result.

Partial order reduction (POR)~\cite{Guo2015, Wachter2013,
  Flanagan2005, Gueta2007, Abdulla2014} identifies equivalent
executions in order to verify a sufficiently large subset. It is
orthogonal to IRS and can be used in conjunction, as we demonstrate via
symbolic traces. IRS differs from other verification techniques that
handle thread scheduling explicitly in ensuring a safe program
execution as soon as a single symbolic trace is found.

In addition to scheduling, a source of non-determinism are relaxed
memory models in modern architectures. In \cite{Burckhardt2008}, a
memory monitoring approach is proposed to make sure that sequential
consistency is maintained during the execution of a program. Fang et
al.\ \cite{Fang2003} present an automated memory fence insertion
technique to enforce SC using instrumentation at the source code
level. In both cases, the program can be safely verified under the
assumption that SC holds with a reduced state space. Similarly to IRS,
these approaches restrict the amount of non-determinism. However, in
contrast to IRS, they are not able to dynamically adapt the amount of
non-determinism and are restricted to non-determinism due to relaxed
memory access. Slightly related, synchronization synthesis, for
example presented by Gupta et al.~\cite{Gupta2015}, automatically
inserts locks and other synchronization primitives that are more
powerful than fences in that also scheduler-related non-determinism
can be eliminated. However, this technique may introduce deadlocks
into a program~\cite{Gupta2015}, hence it is unsuitable for IRS, where
we have to rely on the fact that a verified schedule does not limit
the program's functionality.


\section{Conclusion}
Iteratively Relaxed Scheduling enables to adjust both the amount of
scheduler-related non-determinism and the size of the relevant part of
a program's state space to be verified. This paper discusses issues of
how to efficiently implement IRS in terms of execution time overhead,
i.e., how to efficiently encode and enforce scheduling
constraints. Furthermore, we formalize the requirements on verifiers
for IRS. Support for non-terminating programs is left for future work.

Our experimental results show that iteratively relaxing scheduling
constraints can reduce the execution time overhead. Thereby, we give
evidence that IRS indeed allows to \emph{adjust} both the verification
delay and the incurred execution time overhead in order to find a
sweet spot. Interestingly, we found cases in which a much earlier
reduction of execution time overhead is obtained by choosing favorable
scheduling constraints, which suggests that execution time performance
does not simply rely on the number of scheduling constraints but to a
large extend also on their structure.

For future work, we suggest to investigate the applicability of IRS in
specific use cases. For example, software that is designed
sequentially for safety reasons might be executed concurrently under
IRS. Interesting questions are how many admissible traces necessary to
benefit from concurrency and how long it takes a verifier to check
this number of traces for the programs of the use case. Furthermore, a
sensible automation of finding fast scheduling constraints is of
interest, i.e., a heuristic that finds scheduling constraints with a
structure that induces a low execution time overhead.


\bibliographystyle{plain}
\bibliography{bibliography}

\begin{thebibliography}{10}

\bibitem{Abdulla2014}
Parosh~Aziz Abdulla, Stavros Aronis, Bengt Jonsson, and Konstantinos~F.
  Sagonas.
\newblock Optimal dynamic partial order reduction.
\newblock In Suresh Jagannathan and Peter Sewell, editors, {\em The 41st Annual
  {ACM} {SIGPLAN-SIGACT} Symposium on Principles of Programming Languages,
  {POPL} '14, San Diego, CA, USA, January 20-21, 2014}, pages 373--384. {ACM},
  2014.

\bibitem{Albert2017}
Elvira Albert, Puri Arenas, Maria~Garcia de~la Banda, Miguel
  G{\'{o}}mez{-}Zamalloa, and Peter~J. Stuckey.
\newblock Context-sensitive dynamic partial order reduction.
\newblock In Rupak Majumdar and Viktor Kuncak, editors, {\em Computer Aided
  Verification - 29th International Conference, {CAV} 2017, Heidelberg,
  Germany, July 24-28, 2017, Proceedings, Part {I}}, volume 10426 of {\em
  Lecture Notes in Computer Science}, pages 526--543. Springer, 2017.

\bibitem{Aviram2010}
Amittai Aviram, Shu{-}Chun Weng, Sen Hu, and Bryan Ford.
\newblock Efficient system-enforced deterministic parallelism.
\newblock In Remzi~H. Arpaci{-}Dusseau and Brad Chen, editors, {\em 9th
  {USENIX} Symposium on Operating Systems Design and Implementation, {OSDI}
  2010, October 4-6, 2010, Vancouver, BC, Canada, Proceedings}, pages 193--206.
  {USENIX} Association, 2010.

\bibitem{Bergan2010}
Tom Bergan, Owen Anderson, Joseph Devietti, Luis Ceze, and Dan Grossman.
\newblock Coredet: a compiler and runtime system for deterministic
  multithreaded execution.
\newblock In James~C. Hoe and Vikram~S. Adve, editors, {\em Proceedings of the
  15th International Conference on Architectural Support for Programming
  Languages and Operating Systems, {ASPLOS} 2010, Pittsburgh, Pennsylvania,
  USA, March 13-17, 2010}, pages 53--64. {ACM}, 2010.

\bibitem{Bergan2013}
Tom Bergan, Luis Ceze, and Dan Grossman.
\newblock Input-covering schedules for multithreaded programs.
\newblock In Antony~L. Hosking, Patrick~Th. Eugster, and Cristina~V. Lopes,
  editors, {\em Proceedings of the 2013 {ACM} {SIGPLAN} International
  Conference on Object Oriented Programming Systems Languages {\&}
  Applications, {OOPSLA} 2013, part of {SPLASH} 2013, Indianapolis, IN, USA,
  October 26-31, 2013}, pages 677--692. {ACM}, 2013.

\bibitem{Beyer2012a}
Dirk Beyer, Thomas~A. Henzinger, M.~Erkan Keremoglu, and Philipp Wendler.
\newblock Conditional model checking: a technique to pass information between
  verifiers.
\newblock In Will Tracz, Martin~P. Robillard, and Tevfik Bultan, editors, {\em
  20th {ACM} {SIGSOFT} Symposium on the Foundations of Software Engineering
  (FSE-20), SIGSOFT/FSE'12, Cary, NC, {USA} - November 11 - 16, 2012}, page~57.
  {ACM}, 2012.

\bibitem{Burckhardt2008}
Sebastian Burckhardt and Madanlal Musuvathi.
\newblock Effective program verification for relaxed memory models.
\newblock In Aarti Gupta and Sharad Malik, editors, {\em Computer Aided
  Verification, 20th International Conference, {CAV} 2008, Princeton, NJ, USA,
  July 7-14, 2008, Proceedings}, volume 5123 of {\em Lecture Notes in Computer
  Science}, pages 107--120. Springer, 2008.

\bibitem{Clarke1999}
Edmund~M. Clarke, Orna Grumberg, Marius Minea, and Doron Peled.
\newblock State space reduction using partial order techniques.
\newblock {\em STTT}, 2(3):279--287, 1999.

\bibitem{Cordeiro2011}
Lucas~C. Cordeiro and Bernd Fischer.
\newblock Verifying multi-threaded software using smt-based context-bounded
  model checking.
\newblock In {\em Proceedings of the 33rd International Conference on Software
  Engineering, {ICSE} 2011, Waikiki, Honolulu , HI, USA, May 21-28, 2011},
  pages 331--340, 2011.

\bibitem{Cui2013}
Heming Cui, Jir{\'{\i}} Simsa, Yi{-}Hong Lin, Hao Li, Ben Blum, Xinan Xu,
  Junfeng Yang, Garth~A. Gibson, and Randal~E. Bryant.
\newblock Parrot: a practical runtime for deterministic, stable, and reliable
  threads.
\newblock In Michael Kaminsky and Mike Dahlin, editors, {\em {ACM} {SIGOPS}
  24th Symposium on Operating Systems Principles, {SOSP} '13, Farmington, PA,
  USA, November 3-6, 2013}, pages 388--405. {ACM}, 2013.

\bibitem{Cui2011}
Heming Cui, Jingyue Wu, John Gallagher, Huayang Guo, and Junfeng Yang.
\newblock Efficient deterministic multithreading through schedule relaxation.
\newblock In Ted Wobber and Peter Druschel, editors, {\em Proceedings of the
  23rd {ACM} Symposium on Operating Systems Principles 2011, {SOSP} 2011,
  Cascais, Portugal, October 23-26, 2011}, pages 337--351. {ACM}, 2011.

\bibitem{Fang2003}
Xing Fang, Jaejin Lee, and Samuel~P. Midkiff.
\newblock Automatic fence insertion for shared memory multiprocessing.
\newblock In Utpal Banerjee, Kyle Gallivan, and Antonio Gonz{\'{a}}lez,
  editors, {\em Proceedings of the 17th Annual International Conference on
  Supercomputing, {ICS} 2003, San Francisco, CA, USA, June 23-26, 2003}, pages
  285--294. {ACM}, 2003.

\bibitem{Fischer2013}
Bernd Fischer, Omar Inverso, and Gennaro Parlato.
\newblock Cseq: {A} concurrency pre-processor for sequential {C} verification
  tools.
\newblock In Ewen Denney, Tevfik Bultan, and Andreas Zeller, editors, {\em 2013
  28th {IEEE/ACM} International Conference on Automated Software Engineering,
  {ASE} 2013, Silicon Valley, CA, USA, November 11-15, 2013}, pages 710--713.
  {IEEE}, 2013.

\bibitem{Flanagan2005}
Cormac Flanagan and Patrice Godefroid.
\newblock Dynamic partial-order reduction for model checking software.
\newblock In Jens Palsberg and Mart{\'{\i}}n Abadi, editors, {\em Proceedings
  of the 32nd {ACM} {SIGPLAN-SIGACT} Symposium on Principles of Programming
  Languages, {POPL} 2005, Long Beach, California, USA, January 12-14, 2005},
  pages 110--121. {ACM}, 2005.

\bibitem{Godefroid1996}
Patrice Godefroid.
\newblock {\em Partial-Order Methods for the Verification of Concurrent Systems
  - An Approach to the State-Explosion Problem}, volume 1032 of {\em Lecture
  Notes in Computer Science}.
\newblock Springer, 1996.

\bibitem{Gueta2007}
Guy Gueta, Cormac Flanagan, Eran Yahav, and Mooly Sagiv.
\newblock Cartesian partial-order reduction.
\newblock In Dragan Bosnacki and Stefan Edelkamp, editors, {\em SPIN}, volume
  4595 of {\em Lecture Notes in Computer Science}, pages 95--112. Springer,
  2007.

\bibitem{Guo2015}
Shengjian Guo, Markus Kusano, Chao Wang, Zijiang Yang, and Aarti Gupta.
\newblock Assertion guided symbolic execution of multithreaded programs.
\newblock In Elisabetta~Di Nitto, Mark Harman, and Patrick Heymans, editors,
  {\em Proceedings of the 2015 10th Joint Meeting on Foundations of Software
  Engineering, {ESEC/FSE} 2015, Bergamo, Italy, August 30 - September 4, 2015},
  pages 854--865. {ACM}, 2015.

\bibitem{Gupta2015}
Ashutosh Gupta, Thomas~A. Henzinger, Arjun Radhakrishna, Roopsha Samanta, and
  Thorsten Tarrach.
\newblock Succinct representation of concurrent trace sets.
\newblock In Sriram~K. Rajamani and David Walker, editors, {\em POPL}, pages
  433--444. {ACM}, 2015.

\bibitem{Inverso2014}
Omar Inverso, Ermenegildo Tomasco, Bernd Fischer, Salvatore {La Torre}, and
  Gennaro Parlato.
\newblock Bounded model checking of multi-threaded {C} programs via lazy
  sequentialization.
\newblock In Armin Biere and Roderick Bloem, editors, {\em Computer Aided
  Verification - 26th International Conference, {CAV} 2014}, volume 8559 of
  {\em Lecture Notes in Computer Science}, pages 585--602. Springer, 2014.

\bibitem{Kahkonen2012}
Kari K{\"{a}}hk{\"{o}}nen, Olli Saarikivi, and Keijo Heljanko.
\newblock Using unfoldings in automated testing of multithreaded programs.
\newblock In Michael Goedicke, Tim Menzies, and Motoshi Saeki, editors, {\em
  {IEEE/ACM} International Conference on Automated Software Engineering,
  ASE'12, Essen, Germany, September 3-7, 2012}, pages 150--159. {ACM}, 2012.

\bibitem{Kahkonen2014}
Kari K{\"{a}}hk{\"{o}}nen, Olli Saarikivi, and Keijo Heljanko.
\newblock Unfolding based automated testing of multithreaded programs.
\newblock {\em Automated Software Engineering}, pages 1--41, 2014.

\bibitem{Lal2009}
Akash Lal and Thomas~W. Reps.
\newblock Reducing concurrent analysis under a context bound to sequential
  analysis.
\newblock {\em Formal Methods in System Design}, 35(1):73--97, 2009.

\bibitem{Lauterburg2010}
Steven Lauterburg, Rajesh~K. Karmani, Darko Marinov, and Gul Agha.
\newblock Evaluating ordering heuristics for dynamic partial-order reduction
  techniques.
\newblock In David~S. Rosenblum and Gabriele Taentzer, editors, {\em FASE},
  volume 6013 of {\em Lecture Notes in Computer Science}, pages 308--322.
  Springer, 2010.

\bibitem{Liu2011}
Tongping Liu, Charlie Curtsinger, and Emery~D. Berger.
\newblock Dthreads: efficient deterministic multithreading.
\newblock In Ted Wobber and Peter Druschel, editors, {\em Proceedings of the
  23rd {ACM} Symposium on Operating Systems Principles 2011, {SOSP} 2011},
  pages 327--336. {ACM}, 2011.

\bibitem{Mazurkiewicz1986}
Antoni~W. Mazurkiewicz.
\newblock Trace theory.
\newblock In Wilfried Brauer, Wolfgang Reisig, and Grzegorz Rozenberg, editors,
  {\em Advances in Petri Nets}, volume 255 of {\em Lecture Notes in Computer
  Science}, pages 279--324. Springer, 1986.

\bibitem{McMillan1992}
Kenneth~L. McMillan.
\newblock Using unfoldings to avoid the state explosion problem in the
  verification of asynchronous circuits.
\newblock In Gregor von Bochmann and David~K. Probst, editors, {\em Computer
  Aided Verification, Fourth International Workshop, {CAV} '92, Montreal,
  Canada, June 29 - July 1, 1992, Proceedings}, volume 663 of {\em Lecture
  Notes in Computer Science}, pages 164--177. Springer, 1992.

\bibitem{Metzler2017}
Patrick Metzler, Habib Saissi, P{\'{e}}ter Bokor, and Neeraj Suri.
\newblock Quick verification of concurrent programs by iteratively relaxed
  scheduling.
\newblock In Grigore Rosu, Massimiliano~Di Penta, and Tien~N. Nguyen, editors,
  {\em ASE}, pages 776--781. {IEEE}, 2017.

\bibitem{Metzler2019}
Patrick Metzler, Neeraj Suri, and Georg Weissenbacher.
\newblock Extracting safe thread schedules from incomplete model checking
  results.
\newblock In Fabrizio Biondi, Thomas Given{-}Wilson, and Axel Legay, editors,
  {\em Model Checking Software - 26th International Symposium, {SPIN} 2019,
  Beijing, China, July 15-16, 2019, Proceedings}, volume 11636 of {\em Lecture
  Notes in Computer Science}, pages 153--171. Springer, 2019.

\bibitem{Musuvathi2007}
Madanlal Musuvathi and Shaz Qadeer.
\newblock Iterative context bounding for systematic testing of multithreaded
  programs.
\newblock In Jeanne Ferrante and Kathryn~S. McKinley, editors, {\em Proceedings
  of the {ACM} {SIGPLAN} 2007 Conference on Programming Language Design and
  Implementation, San Diego, California, USA, June 10-13, 2007}, pages
  446--455. {ACM}, 2007.

\bibitem{Nguyen2017}
Truc~L. Nguyen, Peter Schrammel, Bernd Fischer, Salvatore {La Torre}, and
  Gennaro Parlato.
\newblock Parallel bug-finding in concurrent programs via reduced interleaving
  instances.
\newblock In Grigore Rosu, Massimiliano~Di Penta, and Tien~N. Nguyen, editors,
  {\em Proceedings of the 32nd {IEEE/ACM} International Conference on Automated
  Software Engineering, {ASE} 2017, Urbana, IL, USA, October 30 - November 03,
  2017}, pages 753--764. {IEEE} Computer Society, 2017.

\bibitem{Olszewski2009}
Marek Olszewski, Jason Ansel, and Saman~P. Amarasinghe.
\newblock Kendo: efficient deterministic multithreading in software.
\newblock In {\em Proceedings of the 14th International Conference on
  Architectural Support for Programming Languages and Operating Systems,
  {ASPLOS} 2009, Washington, DC, USA, March 7-11, 2009}, pages 97--108, 2009.

\bibitem{Qadeer2005}
Shaz Qadeer and Jakob Rehof.
\newblock Context-bounded model checking of concurrent software.
\newblock In Nicolas Halbwachs and Lenore~D. Zuck, editors, {\em Tools and
  Algorithms for the Construction and Analysis of Systems, 11th International
  Conference, {TACAS} 2005}, volume 3440 of {\em Lecture Notes in Computer
  Science}, pages 93--107. Springer, 2005.

\bibitem{Rabinovitz2005}
Ishai Rabinovitz and Orna Grumberg.
\newblock Bounded model checking of concurrent programs.
\newblock In Kousha Etessami and Sriram~K. Rajamani, editors, {\em Computer
  Aided Verification, 17th International Conference, {CAV} 2005, Edinburgh,
  Scotland, UK, July 6-10, 2005, Proceedings}, volume 3576 of {\em Lecture
  Notes in Computer Science}, pages 82--97. Springer, 2005.

\bibitem{Ramalingam2000}
G.~Ramalingam.
\newblock Context-sensitive synchronization-sensitive analysis is undecidable.
\newblock {\em {ACM} Trans. Program. Lang. Syst.}, 22(2):416--430, 2000.

\bibitem{Rodriguez2015}
C{\'{e}}sar Rodr{\'{\i}}guez, Marcelo Sousa, Subodh Sharma, and Daniel
  Kroening.
\newblock Unfolding-based partial order reduction.
\newblock In Luca Aceto and David de~Frutos{-}Escrig, editors, {\em 26th
  International Conference on Concurrency Theory, {CONCUR} 2015}, volume~42 of
  {\em LIPIcs}, pages 456--469. Schloss Dagstuhl - Leibniz-Zentrum fuer
  Informatik, 2015.

\bibitem{Sousa2017}
Marcelo Sousa, C{\'{e}}sar Rodr{\'{\i}}guez, Vijay D'Silva, and Daniel
  Kroening.
\newblock Abstract interpretation with unfoldings.
\newblock In Rupak Majumdar and Viktor Kuncak, editors, {\em Computer Aided
  Verification - 29th International Conference, {CAV} 2017, Heidelberg,
  Germany, July 24-28, 2017, Proceedings, Part {II}}, volume 10427 of {\em
  Lecture Notes in Computer Science}, pages 197--216. Springer, 2017.

\bibitem{Valmari1996}
Antti Valmari.
\newblock The state explosion problem.
\newblock In Wolfgang Reisig and Grzegorz Rozenberg, editors, {\em Lectures on
  Petri Nets {I:} Basic Models, Advances in Petri Nets, the volumes are based
  on the Advanced Course on Petri Nets, held in Dagstuhl, September 1996},
  volume 1491 of {\em Lecture Notes in Computer Science}, pages 429--528.
  Springer, 1996.

\bibitem{Wachter2013}
Bj{\"{o}}rn Wachter, Daniel Kroening, and Jo{\"{e}}l Ouaknine.
\newblock Verifying multi-threaded software with impact.
\newblock In {\em Formal Methods in Computer-Aided Design, {FMCAD} 2013,
  Portland, OR, USA, October 20-23, 2013}, pages 210--217. {IEEE}, 2013.

\bibitem{LLVM}
The {LLVM} compiler infrastructure.
\newblock \url{https://llvm.org}.

\bibitem{SV-COMP}
Collection of verification tasks.
\newblock \url{https://github.com/dbeyer/sv-benchmarks}.

\end{thebibliography}

\appendix
\section{Experimental Results}\label{app:results}
The following table shows our detailed measurement results. The
columns contain the benchmark name (-opt means with optimized trace
prefixes), the number of constraints in the respective trace prefix,
the mean execution time in \(\mu s\)
and the execution time overhead compared to the uninstrumented
benchmark version.

\center
\begin{supertabular}{lrrr}
  \toprule Benchmark & Constraints & Time & Overhead\%\\
  \toprule

  \midrule
  bigshot & 1 & 124 & 5\%\\
  bigshot & 0 & 121 & 3\%\\
  \midrule dekker & 2 & 115 & 4\%\\
  dekker & 1 & 114 & 3\%\\
  dekker & 0 & 113 & 2\%\\
  \midrule fibonacci & 98 & 176 & 13\%\\
  fibonacci & 44 & 169 & 9\%\\
  fibonacci & 24 & 181 & 12\%\\
  fibonacci & 0 & 166 & 6\%\\
  \midrule lamport & 16 & 123 & 12\%\\
  lamport & 15 & 123 & 12\%\\
  lamport & 10 & 124 & 13\%\\
  lamport & 7 & 124 & 13\%\\
  lamport & 6 & 124 & 13\%\\
  lamport & 4 & 123 & 12\%\\
  lamport & 2 & 123 & 12\%\\
  lamport & 1 & 124 & 13\%\\
  lamport & 0 & 113 & 3\%\\
  \midrule peterson & 28 & 124 & 8\%\\
  peterson & 24 & 125 & 9\%\\
  peterson & 22 & 122 & 6\%\\
  peterson & 1 & 123 & 7\%\\
  peterson & 0 & 113 & -2\%\\
  \midrule shared pointer & 3 & 135 & 22\%\\
  shared pointer & 2 & 134 & 21\%\\
  shared pointer & 1 & 133 & 20\%\\
  shared pointer & 0 & 115 & 4\%\\
  \midrule indexer(15) & 12 & 7538 & 2692\%\\
  indexer(15) & 8 & 7603 & 2716\%\\
  indexer(15) & 4 & 6793 & 2416\%\\
  indexer(15) & 3 & 5412 & 1904\%\\
  indexer(15) & 2 & 435 & 61\%\\
  indexer(15) & 1 & 299 & 11\%\\
  indexer(15) & 0 & 235 & -13\%\\
  \midrule last zero(16) & 15 & 10664 & 4288\%\\
  last zero(16) & 8 & 5286 & 2075\%\\
  last zero(16) & 5 & 492 & 102\%\\
  last zero(16) & 1 & 263 & 8\%\\
  last zero(16) & 0 & 230 & -5\%\\
  \midrule indexer(15)-opt & 12 & 5558 & 2841\%\\
  indexer(15)-opt & 9 & 279 & 48\%\\
  indexer(15)-opt & 6 & 257 & 36\%\\
  indexer(15)-opt & 0 & 215 & 14\%\\
  \midrule last zero(16)-opt & 15 & 378 & 94\%\\
  last zero(16)-opt & 8 & 269 & 38\%\\
  last zero(16)-opt & 5 & 253 & 30\%\\
  last zero(16)-opt & 1 & 250 & 28\%\\
  last zero(16)-opt & 0 & 223 & 14\%\\
  \bottomrule
\end{supertabular}


\end{document}